\documentclass[manuscript]{aastex61}

\usepackage{url}
\usepackage{tabularx}
\usepackage{amsmath}
\usepackage{rotating}
\usepackage{multirow}
\usepackage{xcolor}

\setcitestyle{citesep={,}}

\shorttitle{SDO/HMI LOS observables for electron beam heating models}
\shortauthors{Sadykov et al.}

\begin{document}
	
\title{Response of SDO/HMI observables to heating of the solar atmosphere by precipitating high-energy electrons}
\email{sadykov@baeri.org}

\author{Viacheslav M. Sadykov}
\affiliation{NASA Ames Research Center, Moffett Field, CA 94035, USA}
\affiliation{Bay Area Environmental Research Institute, Moffett Field, CA 94035, USA}

\author{Alexander G. Kosovichev}
\affiliation{Center for Computational Heliophysics, New Jersey Institute of Technology, Newark, NJ 07102, USA}
\affiliation{Department of Physics, New Jersey Institute of Technology, Newark, NJ 07102, USA}
\affiliation{NASA Ames Research Center, Moffett Field, CA 94035, USA}

\author{Irina N. Kitiashvili}
\affiliation{NASA Ames Research Center, Moffett Field, CA 94035, USA}
\affiliation{Bay Area Environmental Research Institute, Moffett Field, CA 94035, USA}

\author{Graham S. Kerr}
\affiliation{NASA Goddard Space Flight Center, Heliophysics Sciences Division, Code 671, Greenbelt MD 20771, USA}

\begin{abstract}
	We perform an analysis of the line-of-sight (LOS) observables of the Helioseismic and Magnetic Imager (HMI) onboard the Solar Dynamics Observatory (SDO) for models of the solar atmosphere heated by precipitating high-energy electrons during solar flares. The radiative hydrodynamic (RADYN) flare models are obtained from the F-CHROMA database. The Stokes profiles for the Fe\,6173\,{\AA} line observed by SDO/HMI are calculated using the radiative transfer code RH1.5D assuming statistical equilibrium for atomic level populations and imposing uniform background vertical magnetic field of various strength. The SDO/HMI observing sequence and LOS data processing pipeline algorithm are applied to derive the observables (continuum intensity, line depth, Doppler velocity, LOS magnetic field). Our results reveal that the strongest deviations of the observables from the actual spectroscopic line parameters are found for the model with a total energy deposited of $E_{total}=1.0\times{}10^{12}$\,erg\,cm$^{-2}$, injected with power-law spectral index of $\delta=3$ above a low-energy cutoff of $E_{c}=25$\,keV. The magnitudes of the velocity and magnetic field deviations depend on the imposed magnetic field, and can reach 0.35\,km/s for LOS velocities, 90\,G for LOS magnetic field, and 3\% for continuum enhancement for the 1000\,G imposed LOS magnetic field setup. For $E_{total}\geq{}3.0\times{}10^{11}$\,erg\,cm$^{-2}$ models, the velocity and magnetic field deviations are most strongly correlated with the energy flux carried by $\sim$50\,keV electrons, and the continuum enhancement is correlated with the synthesized $\sim$55-60\,keV hard X-ray photon flux. The relatively low magnitudes of perturbations of the observables and absence of magnetic field sign reversals suggest that the considered radiative hydrodynamic beam heating models augmented with the uniform vertical magnetic field setups cannot explain strong transient changes found in the SDO/HMI observations.
\end{abstract}


\section{Introduction}
\label{Section:introduction}

	Changes of the solar atmospheric conditions and emission properties during solar flares (e.g. measured magnetic field, plasma flows, and enhancement of spectral line and continuum emission) are of special interest because of their close relation to flare energy deposit and transport properties. Such changes are especially intriguing and hard to explain if observed at the solar photosphere. Even though photospheric variations are often reported in the literature \citep[e.g.][]{Liu17a,Sun17a,Castellanos18a,Song18a}, the question of how the deep layers of the solar atmosphere are strongly perturbed during solar flares is still debated.
	
	The line-of-sight (LOS) observables obtained by the Helioseismic and Magnetic Imager onboard the Solar Dynamics Observatory \citep[SDO/HMI,][]{Scherrer12a,Couvidat12b} currently represent one of the most widely-used data products in solar physics. To obtain the LOS observables, HMI images the Fe\,I\,6173\,{\AA} line, over six wavelength points (filtergrams) in two polarizations (right-circular, RCP, and left-circular, LCP). In quiet-Sun conditions, the Fe\,I\,6173\,{\AA} line forms in the photospheric height range of 0-300\,km above the $\tau{}_{5000}=1$ \citep{Norton06a,Nagashima14a,Kitiashvili15a}, making the SDO/HMI observables sensitive to the perturbations of the deep atmospheric layers. During solar flares, the Fe\,I\,6173\,{\AA} line continuum may form higher, at $\sim$200\,km above the $\tau{}_{5000}=1$ photosphere \citep{MartinezOliveiros12a}, and may even include a contribution from the chromospheric heights \citep{Heinzel17a}. Correct interpretation of these measurements is important for understanding the underlying physics and expansion of our knowledge about the magnetic energy release and photospheric impacts of solar flares.
	
	The HMI observables (line depth and width, continuum intensity, Doppler velocity, and LOS magnetic field) are calculated from the filtergrams using a Gaussian line-profile model~\citep{Couvidat16a}. It takes 45\,s to scan the full line. If the Fe\,I\,6173\,{\AA} line profile and continuum vary faster than or comparable to the HMI LOS cadence, then one should expect deviations of the HMI observables from the actual properties of the line profile (and consequently, the retrieved properties of the atmosphere). For example, Swedish Solar Telescope (SST) observations of C\,2.0 class flare event revealed variations of the optical continuum near H$\alpha{}$ line to be as fast as 9\,s \citep{Jess08a}. If such fast variations take place for the continuum near Fe\,I\,6173\,{\AA} line, this cannot be captured by HMI pipeline. Taking into account the non-instantaneous nature of the measurements is especially important for interpretation of ``magnetic transients''~--- reversible sharp changes of magnetic field measurements during solar flares \citep{Zirin81a,Patterson84a}. Previous reports of such magnetic transients, often accompanied by magnetic polarity reversals observed from SOHO/MDI and SDO/HMI, concluded that these transients may represent real changes of the magnetic field strength \citep{Kosovichev01a,Zharkova02a,Harker13a,Lozitsky19} or may be artifacts due to the data analysis algorithms \citep{Qiu03a,Mravcova17a,Maurya12a}. The correctness of the measured measured LOS velocities \citep{Maurya12a} is also under debate.
	
	The nature of the observed enhancements of the continuum intensity near Fe\,I\,6173\,{\AA} line is also not fully understood. \citet{Svanda18a} analyzed the simultaneous observations by HMI and Hinode Solar Optical Telescope Spectro-polarimeter \citep[Hinode SOT/SP,][]{Kosugi07a,Tsuneta08a} and found that the pseudocontinuum derived from HMI measurements does not approximate the continuum value properly. \citet{Mravcova17a} also pointed out that the enhanced HMI continuum might be an artifact of the procedure for calculation of the observables.	In general, origin of the optical continuum emission and its enhancement remains a fundamental open question in our knowledge of solar flares.

	For interpretation of the observed variations, it is important to compare the observational results with the predictions of state-of-the-art models of flare heating. One of the most accepted mechanisms is precipitation of non-thermal electrons accelerated in the corona into the lower layers of the solar atmosphere. Radiative hydrodynamic simulations of the electron beam-heated flare model developed in recent years allow us to calculate response of the photospheric spectral lines and compare it with observations. Currently one of the most advanced codes for flare modeling is RADYN, a radiative hydrodynamic code \citep{Carlsson97a,Abbett99a,Allred05a,Allred06a,Allred15a}. A grid of RADYN models with various parameters of the injected electron energy beams is available online from the F-CHROMA project (\url{http://www.fchroma.org/}).
		
	The precipitating electrons generate bremsstrahlung emission evident in Hard X-ray (HXR) observations. Previously many works \citep[e.g.][]{Qiu03a,Chen05a,Burtseva15a,Huang16a,Kuhar16a,Lee17a,MartinezOliveiros12a,Sharykin17a,Sharykin18b} found that impulsive variations observed in the SDO/HMI measurements are correlated with the HXR signals, both temporarily and spatially, suggesting that the impact felt by the deep photospheric layers are caused by precipitating high-energy electrons. However, there is typically insufficient power carried by electrons with the energy high enough to penetrate through to the photosphere, though accelerated ions or radiative backwarming are among possible explanations \citep[e.g. see discussions in the context of white light flares in][]{Neidig89a,Kerr14a}. Here we investigate predictions of the standard model of non-thermal electron beam in terms of the photospheric response measured by HMI, and compare these predictions with observational constraints.
	
	We use the RADYN models augmented with the uniform vertical magnetic field to simulate the Fe\,I\,6173\,{\AA} line Stokes profiles and derive the corresponding SDO/HMI observables by applying the synthetic data analysis algorithms implemented in the LOS SDO/HMI Joint Science Operations Center (JSOC) pipeline. We analyze deviations of the synthetic observables from the actual line-profile properties and atmospheric conditions in the flare models. The modeling of the Fe\,6173\,{\AA} spectral line and the procedure of SDO/HMI LOS observable calculations are explained in Section~\ref{Section:modeling}. The results are presented in Section~\ref{Section:results}, followed by a discussion in Section~\ref{Section:discussion}.

\section{Modeling of SDO/HMI observables}
\label{Section:modeling}

	\subsection{Calculation of Fe\,I\,6173\,{\AA} Stokes profiles for RADYN flare models.}
	
\begin{figure}[t!]
	\centering
	\includegraphics[width=1.0\linewidth]{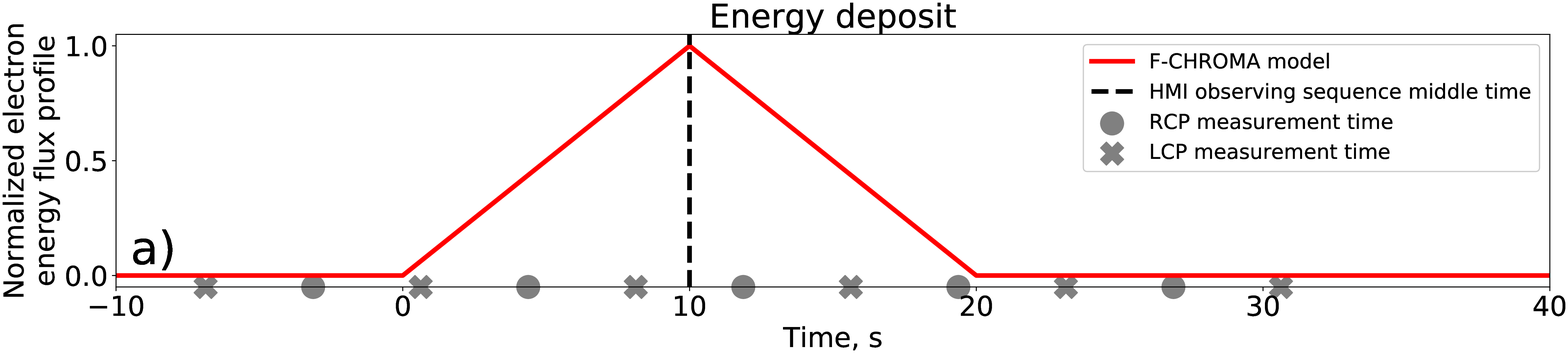} \\
	\includegraphics[width=1.0\linewidth]{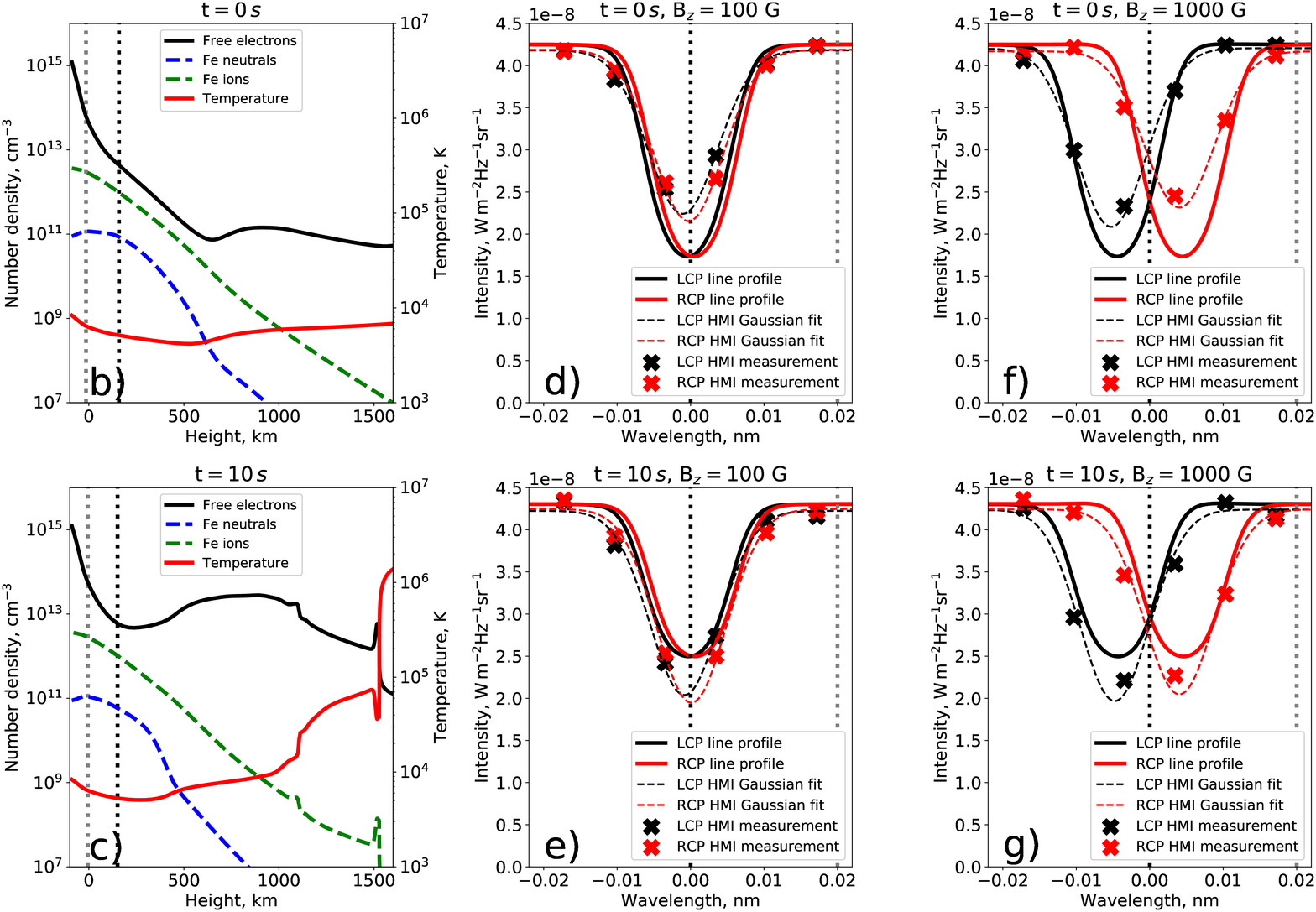}
	\caption{Illustration of (a) normalized electron energy flux profile, (b,c) the atmospheric properties, and (d-g) Fe\,I\,6173\,{\AA} RCP and LCP profiles for $t=0$\,s and $t=10$\,s snapshots of the ``val3c\_d3\_1.0e12\_t20s\_25keV'' RADYN model ($E_{total}=1.0\times{}10^{12}$\,erg\,cm$^{-2}$, $E_{c}=$25\,keV, $\triangle{}t=$20\,s, $\delta{}=$3). Corresponding SDO/HMI filtergram signals calculated for measurement series centered at $t=0$\,s and $t=10$\,s, and their Gaussian fits, are presented in panels (d,e) for 100\,G and in panels (f,g) for 1000\,G vertical uniform magnetic fields. Markers on the x-axis in panel (a) illustrate the HMI LOS observing sequence for RCP (right circular polarization) and LCP (left circular polarization) filtergrams. Dashed black vertical line shows the middle time of the illustrated observing sequence. The dashed vertical lines in panels (b,c) correspond to $\tau{}=1$ optical depths for the Fe\,I\,6173\,{\AA} line center (black) and continuum (gray). The wavelengths representing the line center and continuum are marked by black and gray dashed vertical lines correspondingly in panels d-g.}
	\label{figure1}
\end{figure}

		The F-CHROMA database is a collection of 1D radiative hydrodynamic (RADYN) models of solar flares driven by an electron beam with a power-law electron energy distribution (averaged energy fluxes from 1.5$\times$10$^{9}$\,erg\,cm$^{-2}$s$^{-1}$ to 5.0$\times$10$^{10}$\,erg\,cm$^{-2}$s$^{-1}$, low-energy cutoff values of 10\,keV, 15\,keV, 20\,keV, or 25\,keV, and spectral indexes ranging from 3 to 8) heating the atmosphere for 20\,s. The RADYN code solves the coupled, non-linear, equations of hydrodynamics, radiation transport, and non-equilibrium atomic level populations, using an adaptive 1D vertical grid. The elements that are important for the chromospheric energy balance are treated in non-Local Thermodynamic Equilibrium (NLTE), and other species are included in the radiative loss function in the LTE approximation. The atomic level population and radiation transport equations are solved for 6-level-with-continuum hydrogen, 9-level-with-continuum helium, and 6-level-with-continuum Ca\,II atomic models. For a detailed description of the RADYN code see \citet{Allred15a} and references therein. In the F-CHROMA database, the 1D flare models are calculated with 300 height grid points and 201 frequency points of the radiation spectrum. The initial atmosphere is similar to the VAL3C model \citep{Vernazza81a} but with a somewhat deeper transition region. The temporal profile of the deposited energy flux rate is a triangle; the electron beam heating lasts for 20\,s with a peak at 10\,s (red line in Figure~\ref{figure1}a). Figure~\ref{figure1}b also displays the atmospheric stratification for the pre-flare atmospheres, along with the temporal profiles of energy injection.
	
		For 80 available F-CHROMA models, we calculate the Stokes profiles for the Fe\,I\,6173\,{\AA} line using the RH1.5D code \citep{Pereira15a}~--- the latest massively-parallel version of the RH code \citep{Rybicki91a,Rybicki92a,Uitenbroek01a}. Snapshots of the RADYN flare atmospheres for sequential moments of time are used as input to the RH1.5D code. Since RH1.5D is a stationary code, the NLTE atomic level populations are solved using statistical equilibrium, meaning that non-equilibrium effects are not included in our model. This is somewhat mitigated by using the non-equilibrium electron density from the RADYN models. Such a procedure has been used by others \citep[e.g.,][]{Kerr16a,RDC17a,Sadykov19a}. To take into account magnetic field effects we make the assumption that the beam heating occurs in a vertical flux tube of uniform vertical magnetic field. In this work, we focus on 100\,G and 1000\,G magnetic field setups, and additionally consider 50\,G, 250\,G, 500\,G, and 1500\,G magnetic filed setups for the model ``val3c\_d3\_1.0e12\_t20s\_25keV'' with the strongest photospheric perturbations (as discussed further). The vertical magnetic field is only used for the Fe\,I\,6173\,{\AA} line forward modelling and has no impact on the hydrodynamics. The Fe atomic states (described by 31 Fe\,I bound states, one singly-ionized Fe\,II state, and one twice-ionized Fe\,III state) are calculated assuming non-LTE statistical equilibrium along with the H atom states. Other species (He, O, C, N, Mg, Si, S, Al, Ca, Na, Ni) are calculated in LTE. Experiments showed that non-LTE effects for these species do not affect the Fe\,I\,6173\,{\AA} line profile. We include 2\,km/s non-thermal line broadening to account for microturbulence in the line profile calculations. For each model, RADYN atmospheres were processed through RH code at a 1\,s cadence, and the resulting Stokes profiles are interpolated linearly if needed for calculating the observables. The right-circular polarization (RCP) and left-circular polarization (LCP) signals are derived from the Stokes I and V profiles.
		
		Examples of atmospheric properties and the polarization profiles for the F-CHROMA RADYN model ``val3c\_d3\_1.0e12\_t20s\_25keV'' are illustrated in Figure~\ref{figure1}b-g for $t=0$\,s and $t=10$\,s for 100\,G and 1000\,G uniform vertical magnetic fields. This model has an average deposited energy flux $F=5.0\times{}10^{10}$\,erg\,cm$^{-2}$\,s$^{-1}$ (i.e. $E_{total}=1.0\times{}10^{12}$\,erg\,cm$^{-2}$), a power law index of the injected electron spectrum of $\delta{}=$3, and a low-energy cutoff of $E_{c}=$25\,keV, and demonstrates the strongest disagreement between the modeled LOS observables and the actual line and atmospheric properties among the analyzed models. As demonstrated in Figures~\ref{figure1}d-g, there are no visible qualitative differences between the circular polarization profiles (RCP or LCP) for $t=0$\,s and $t=10$\,s besides the reduced line depth during the peak of the energy deposit. Figure~\ref{figure1}b,c also illustrates the neutral and ionized Fe number densities calculated by the RH1.5D code.
		
	\subsection{Extension of the modeling results.}
		
		Each hydrodynamic flare run lasts 50\,s from the beginning of the beam impact. To process the synthetic results through the SDO/HMI pipeline, we assume that the pre-flare and post-flare states are unvarying and extend the models for 42\,s in both directions. The timing of the flare phases are 1) pre-flare $t=-42-0$\,s, 2) energy injection $t=0-20$\,s, 3) dynamic cooling phase $t=20-50$\,s, and 4) fixed cooling phase $t=50-92$\,s. We are aware that the atmosphere continues to radiate and conduct in dynamic fashion after the flare but assume that this happens much more slowly than the variations during the impulsive phase. For the purposes of this experiment this seems a reasonable approximation.
		
	\subsection{Calculation of line profile properties and SDO/HMI LOS observables.}
		
		For each considered snapshot, we derive the following Fe\,I\,6173\,{\AA} line parameters: line continuum as an averaged intensity at $\pm0.20$\,{\AA} from the line reference wavelength ($\lambda{}_{ref}=$6173.3390\,{\AA}); line depth calculated as the continuum intensity minus the average of the smallest intensities in the LCP and RCP signals; Doppler shift calculated using the center-of-gravity approach $\triangle{}\lambda{} = <\lambda{}>-\lambda{}_{ref} = \int_{\lambda{}_{ref}-0.2\,{\AA}}^{\lambda{}_{ref}+0.2\,{\AA}}(I-I_{c})d\lambda{} - \lambda{}_{ref}$.
		
		To model SDO/HMI LOS observables, we simulate the HMI 45-sec observing sequence illustrated in Figure~\ref{figure1}a. The filtergrams are calculated for $\pm{}34.4\,m{\AA},\pm{}103.2\,m{\AA},$ and $172.0\pm{}\,m{\AA}$ relative to $\lambda{}_{ref}$ \citep{Nagashima14a}, and are linearly interpolated to the observing sequence center time. The temporal order of scanned wavelengths is assumed as in Table 3 of \citet{Schou12a}. The SDO/HMI transmission profiles for each measurement are modeled using the Gaussian profile with $FWHM=76\,m{\AA}$ \citep{Couvidat12a}. Because the typical HMI exposure times are much shorter than 1\,s \citep[about 140\,ms,][]{Couvidat16a} we assume that the filtergrams are taken instantaneously. Examples of the synthesized SDO/HMI measurements for the RADYN model ``val3c\_d3\_1.0e12\_t20s\_25keV'' and the observing sequences centered at $t=0$\,s and $t=10$\,s time moments are illustrated in Figures~\ref{figure1}d-g, together with the Gaussian line fitting assumed by the SDO/HMI LOS pipeline.
		
		We calculate the observables that the HMI pipeline delivers, for comparison to the equivalent properties of the actual spectral lines. Those are the line width, line depth, Doppler shift, and vertical LOS magnetic field, following the procedure described by \citet{Couvidat12b,Couvidat12a,Couvidat16a}. First, we estimate the first and second Fourier components of the line profile separately for each polarization sequence as:
	
		\begin{gather}
		\label{eq:fouriercomponents}
		a_{k}\approx{}\dfrac{2}{6}\sum_{j=0}^{5}I_{j}\textrm{cos}(2k\pi{}\dfrac{2.5-j}{6}), k={1,2} \\
		b_{k}\approx{}\dfrac{2}{6}\sum_{j=0}^{5}I_{j}\textrm{sin}(2k\pi{}\dfrac{2.5-j}{6}), k={1,2}
		\end{gather}
		
		Then, we estimate the line depth, $I_{d}$, the continuum intensity, $I_{c}$, Doppler velocity, $v$, and LOS magnetic field strength, $B$, as:
		
		\begin{gather}
		\label{eq:velocities}
		v_{1} = \dfrac{dv}{d\lambda}\dfrac{T}{2\pi{}}\textrm{atan}\left(\dfrac{b_{1}}{a_{1}}\right) \\
		\lambda{}_{0} = \lambda{}_{ref}+v_{1}\dfrac{d\lambda}{dv} \\
		v = \dfrac{v_{1}^{LCP}+v_{1}^{RCP}}{2} \\
		B = (v_{1}^{LCP}-v_{1}^{RCP})K_{m} \\
		I_{d} = \dfrac{T}{2\sigma{}\sqrt{\pi}}\sqrt{a_{1}^{2}+b_{1}^{2}}\textrm{exp}\left(\dfrac{\pi{}^{2}\sigma{}^{2}}{T^{2}}\right) \\
		\label{eq:Ic}
		I_{c} = \dfrac{1}{6}\sum_{j=0}^{5}\left[I_{j}+I_{d}\textrm{exp}\left(-\dfrac{(\lambda{}_{j}-\lambda{}_{0})^{2}}{\sigma{}^{2}}\right)\right]
		\end{gather}
		
		Here $K_{m}=0.231$\,G\,m$^{-1}$, $\dfrac{dv}{d\lambda} = 48.5624$\,km$\cdot{}$s$^{-1}${\AA}$^{-1}$, $T=$412.8\,m{\AA}. In the SDO/HMI algorithm, a significant error comes from inaccurate determination of the Gaussian line width because of the coarse sampling of the line profile. The correction implemented in the SDO/HMI pipeline is based on the azimuthal average of the width measured at the solar disc center during a period of low solar activity~\citep{Couvidat16a}. In our calculations, we assume that the line width is derived from the preflare state ($\sigma{}=0.0671\,{\AA}$ at $t=0$\,s). In addition, we multiply the line width and the line depth by the correction coefficients, $K_{2}=6/5$ and $K_{1}=5/6$, respectively, as suggested by \citet{Couvidat16a}. Such a correction leads to the closest match between observables and line profile properties during the pre-flare phase.
		
		For each model we calculate three parameters quantifying perturbations of the SDO/HMI observables during the run. The first parameter is the strongest deviation of the velocity observable from the corresponding derivative of the instantaneous line profile during the run. The second parameter is the strongest deviation of the magnetic field observable from its imposed value. The third parameter is the enhancement of the continuum intensity observable defined as $I_{max(flare)}/I_{pre-flare} - 1$, where $I_{max(flare)}$ is the maximum value of the continuum intensity observable during the run, and $I_{pre-flare}$ is its pre-flare (unperturbed) value.

\section{Results}
\label{Section:results}

	In this section, we present a detailed analysis of the Fe\,6173\,{\AA} line profile properties and corresponding HMI observables for one F-CHROMA RADYN model ``val3c\_d3\_1.0e12\_t20s\_25keV'', with $E_{total}=1.0\times{}10^{12}$\,erg\,cm$^{-2}$, $E_{c}=$25\,keV, $\delta{}=$3. We also present the analysis of the strongest perturbations of the HMI observables and the corresponding atmospheric properties, for various values of the total energy flux. In addition, we analyze correlations of the perturbations with the energy flux carried by electrons at and above certain energies, as well as the synthesized Hard X-Ray (HXR) photon fluxes at certain energies.
	
\begin{figure}[t!]
	\centering
	\includegraphics[width=0.9\linewidth]{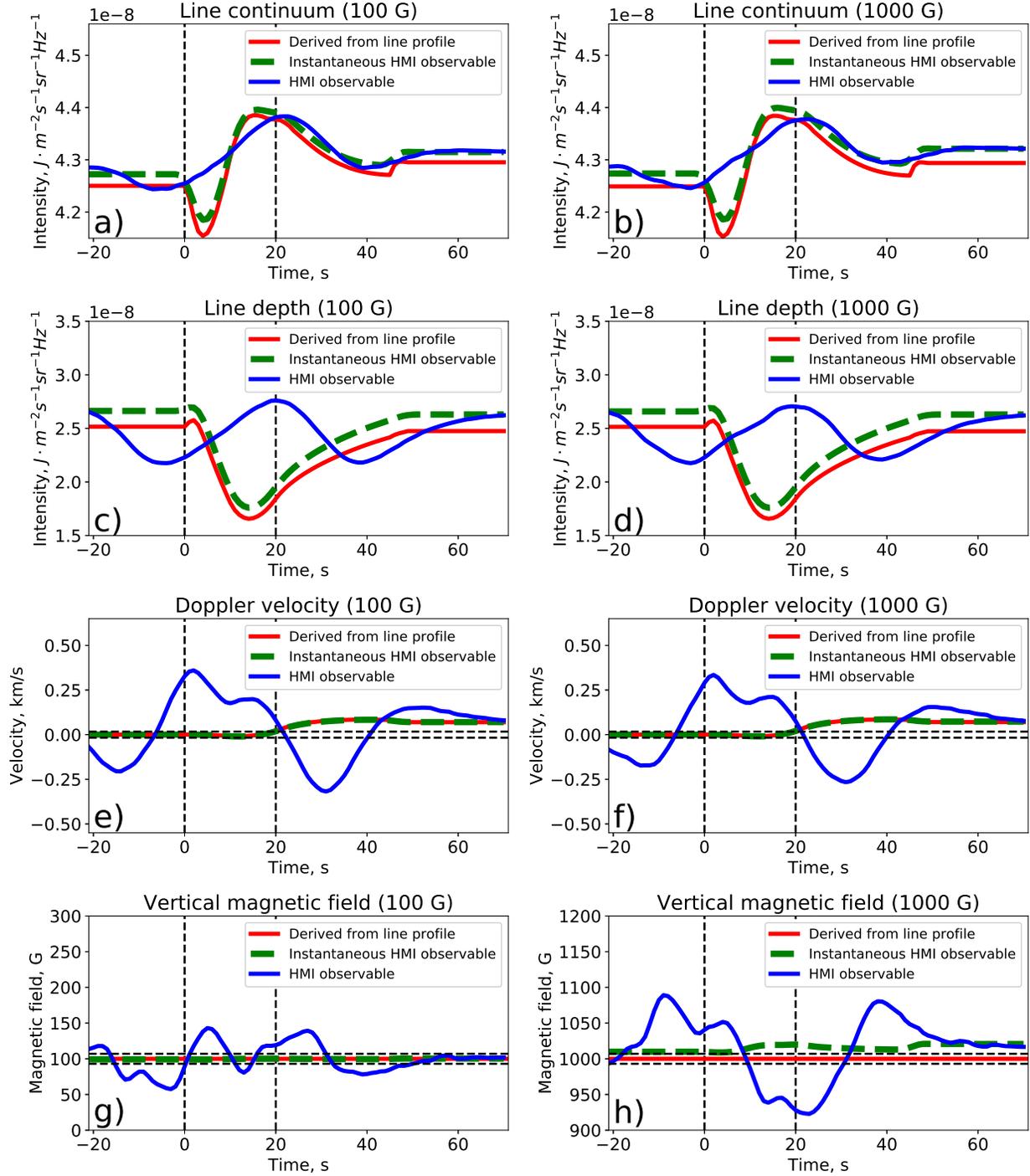}
	\caption{Fe\,I\,6173\,{\AA} line parameters and corresponding simulated SDO/HMI observables for RADYN model ``val3c\_d3\_1.0e12\_t20s\_25keV'' for the vertical uniform 100\,G (panels a, c, e, g) and 1000\,G (panels b, d, f, h) fields. Red curves correspond to the measurements obtained from the native line profiles. Green dashed curves show ``instantaneous'' observables obtained with HMI algorithm applied to the line profile instantaneously. Blue curves show the observables obtained with the HMI algorithm applied with the actual observing sequence timing centered at the reference time. Black dashed vertical lines mark the flare heating phase of the run. Black dashed horizontal lines mark uncertainties of the HMI observables at the disc center caused by the photon noise \citep{Couvidat16a}.}
	\label{figure2}
\end{figure}

	Figure~\ref{figure2} illustrates the Fe\,I\,6173\,{\AA} line properties derived from the simulated line profiles, ``instantaneous'' HMI observables (the results of application of Eqs.~\ref{eq:fouriercomponents}-\ref{eq:Ic} on polarization signals $I_j$ obtained at the same time moment), and the HMI observables obtained with the actual observing sequence timing for the flare model ``val3c\_d3\_1.0e12\_t20s\_25keV''. Two setups with vertical uniform magnetic fields of 100\,G and 1000\,G are shown. Figures~\ref{figure2}a-b show that perturbations of the measured continuum level do not exceed 3\% during the flare. Deviations of the HMI line-depth observable from the values derived from the line profile (Figures~\ref{figure2}c-d) are significant during the heating phase. For example, the line profile depth significantly decreases in the middle of the heating phase, but the corresponding value of the HMI observable centered at this time moment shows an increase. Interestingly, the $\tau{}=1$ heights of the Fe\,I\,6173\,{\AA} line core and continuum do not change strongly; the lines continuum $\tau{}=1$ height increases from -15\,km to -5\,km (relative to the quiet Sun photospheric level), and the $\tau{}=1$ height of the line core decreases from 159\,km to 151\,km, as shown in Figure~\ref{figure1}b,c.
	
	While the instantaneous observables for the Doppler velocity and magnetic field agree with the properties of the line profile, the HMI observables calculated for the time-dependent observing sequence are in strong disagreement with the actual line properties under flaring conditions (Figures~\ref{figure2}e-h). The strongest deviations are found for the Doppler velocities: while the actual values are about 0.1\,km/s, the HMI observable can reach $\pm$ 0.4\,km/s. The magnetic field observable can deviate by 43\,G for the 100\,G background vertical field, and by 89\,G for the 1000\,G. These deviations result from changes of the Fe\,6173\,{\AA} line depth during a flare and the non-instantaneous nature of the HMI observing sequence. We note here that such deviations are higher than the uncertainties of SDO/HMI observables caused by the photon noise at the disc center~\citep[17\,m/s and 7\,G correspondingly,][]{Couvidat16a}. We do not estimate other potential sources of instrumental uncertainties in this study.
	
	As seen in Figure~\ref{figure2}, the deviations depend on the imposed vertical magnetic field. Figure~\ref{figure3} illustrates deviations of the velocity and magnetic field observables from the actual parameters as a function of the background vertical magnetic field strength for model ``val3c\_d3\_1.0e12\_t20s\_25keV''. Interestingly, the stronger magnetic field makes the velocity deviations smaller. This effect, most probably, illustrates that the strength of the Zeeman splitting between RCP and LCP polarization components of the Fe\,I\,6173\,{\AA} line together with the selection of wavelengths for filtergrams affects the derived observables. Also, the measured LOS magnetic field observable becomes close to zero for the 50\,G external magnetic field, yet not resulting in an artificial magnetic field sign reversal. The magnetic field deviations grow with the imposed magnetic field, although the relative changes become smaller, and cannot explain the observed polarity reversals. In fact, the polarity reversals are never observed for the considered flare models.
	
	Figure~\ref{figure4} illustrates the strongest deviations of the Doppler shift and magnetic field as a function of the deposited energy flux. For the illustration, the results are presented only for the high-energy electron beam spectra with power law indexes of 3 (Figure~\ref{figure3}a-d) and low energy cutoff of 25\,keV (panels e-h). As one can see, the deviations depend on the deposited energy flux, and grow on average with the flux value (although showing dependence on other beam parameters). The deviations significantly reduce if $F<5.0\times{}10^{10}$ erg\,cm$^{-2}$s$^{-1}$ (i.e. $E_{total}<1.0\times{}10^{12}$\,erg\,cm$^{-2}$).
	
	The response of the atmosphere depends not only on the energy flux, $F$, but also on the distribution of non-thermal electrons, which is determined by $F$ and two additional parameters, $E_{c}$ and $\delta{}$. Here we analyze correlations between the strongest perturbations of the observables and the energy flux deposited by non-thermal electrons above a certain energy threshold, $F(E>E_{thres})$, and at a certain value, $F(E=E_{el})$. These quantities are determined as:
	
	\begin{gather}
	\label{eq:ethres}
	F(E>E_{thres}) = F\left(\dfrac{E_{thres}}{E_{c}}\right)^{-\delta{}+2}, ~~~E_{thres}\geq{}E_{c} \\
	F(E>E_{thres}) = F, ~~~E_{thres}<E_{c} \\
	F(E=E_{el}) = F\dfrac{\delta{}-2}{E_{c}}\left(\dfrac{E_{el}}{E_{c}}\right)^{-\delta{}+1}, ~~~E_{el}\geq{}E_{c}
	\end{gather}
	
	We also synthesize the Hard X-Ray (HXR) photon flux at a certain energy, $F_{ph}(E=E_{ph})$, following \citet{Brown71a}, which is related to the $F(E=E_{el})$ as:
	
	\begin{gather}
	\label{eq:ethres_ph}
	F_{ph}(E=E_{ph})\sim{}F(E=E_{el})\dfrac{\delta{}-2}{\delta{}-1}
	\end{gather}
		
	To analyze the correlations we calculate non-parametric Kendall's $\tau$ coefficient (Kendall's rank correlation coefficient) defined as:
		
	\begin{gather}
	\label{eq:Kendallstau}
	\tau{} = \dfrac{2}{n(n-1)}\sum_{i<j}sgn(x_{i}-x_{j})sgn(y_{i}-y_{j})
	\end{gather}
	
	Here $\{x_{i}\}$ and $\{y_{i}\}$ are the values of the considered pair of parameter; $sgn$ is a sign operator; $n$ is a number of elements in each data set. Kendall's $\tau$ ranges between -1 and 1, and its value is expected to be 0 for independent data sets. We calculate a p-value for a hypothesis test whose null hypothesis is an absence of association ($\tau{}=0$). Low p-value ($<$ 0.05) indicates that the data points are  aligned better with the presence of correlation hypothesis with respect to the absence of correlation hypothesis ($\tau{}=0$).
	
	The Kendall's $\tau$ correlation coefficient as a function of $E_{thres}$, $E_{el}$, and $E_{ph}$, is presented in Figure~\ref{figure5} for the models of $F\ge{}1.5\times{}10^{10}$\,erg\,cm$^{-2}$s$^{-1}$. The x-axes in panels (d-i) are restricted to lower limit of 25\,keV which is the highest low-energy cutoff value among the considered RADYN models. All presented correlations are statistically significant except the high-energy regions marked by red in Figures~\ref{figure5}c~and~\ref{figure5}f. As evident from Figures~\ref{figure5}a~and~\ref{figure5}d, the velocity deviations depend on the electrons of 15-20\,keV and above, and have the strongest correlation with the $\sim$50\,kev electron energy flux. The same is true for the magnetic field deviations (Figures~\ref{figure5}b~and~~\ref{figure5}e) except that the correlations are the strongest for the $\geq$15\,keV electron flux. Figures~\ref{figure5}c~and~\ref{figure5}f demonstrate that the continuum intensity enhancement depends on the entire non-thermal electron spectra, with the strongest contribution from $\sim$35\,keV electrons. The velocity and magnetic field deviations also correlate with the HXR photon flux. The correlations are higher for higher HXR photon energies (Figures~\ref{figure5}gh), with no peak for a particular energy. Contrary, the continuum intensity enhancement correlations with the HXR photon flux clearly peak at $\sim$55-60\,keV.

\section{Discussion and Conclusion}
\label{Section:discussion}

\begin{figure}[t!]
	\centering
	\includegraphics[width=1.0\linewidth]{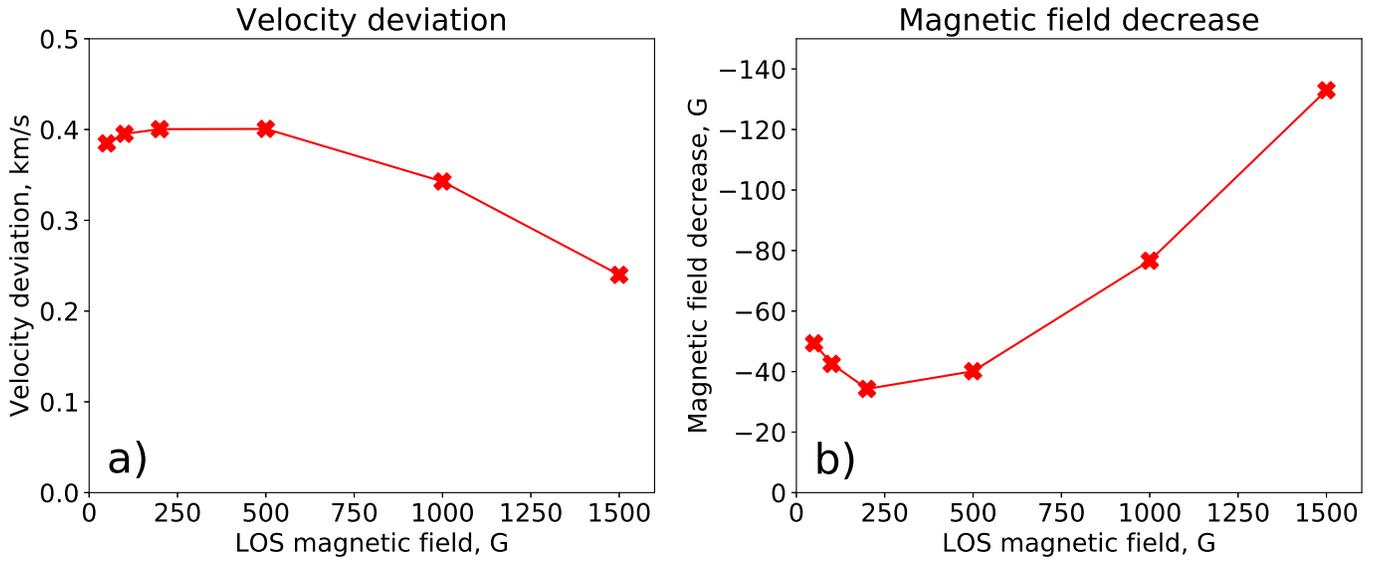}
	\caption{The strongest deviations of the HMI observables as a function of the LOS magnetic field strength for the RADYN model ``val3c\_d3\_1.0e12\_t20s\_25keV''; (a) the LOS velocity observable with respect to the Fe\,I\,6173\,{\AA} Doppler shift; and (b) the LOS magnetic field observable with respect to the uniform vertical magnetic field.}
	\label{figure3}
\end{figure}

\begin{figure}[t!]
	\centering
	\includegraphics[width=0.9\linewidth]{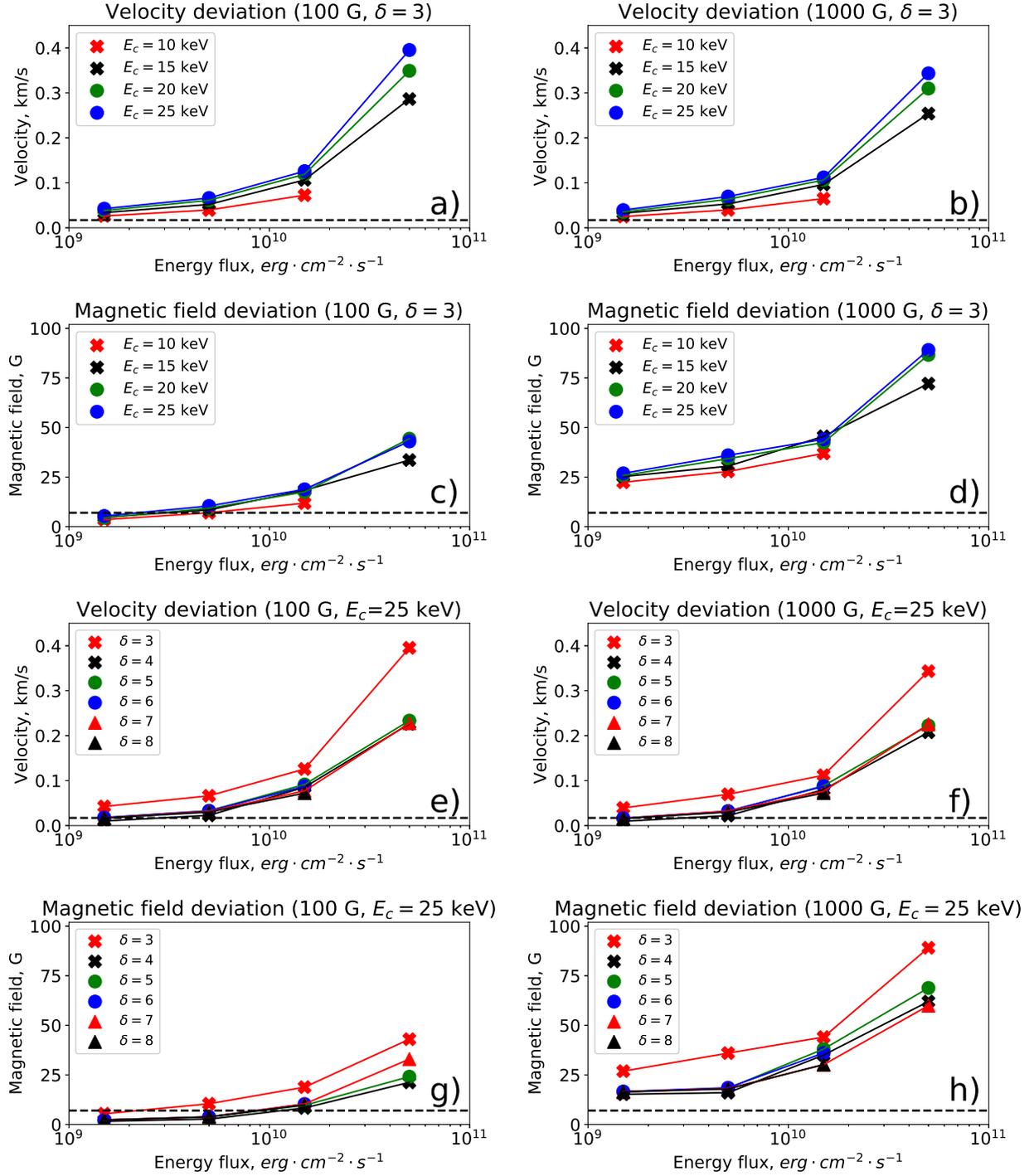}
	\caption{Illustration of the strongest deviations of Fe\,I\,6173\,{\AA} line parameters from the simulated HMI observables for various flare models in the presence of 100\,G (panels a, c, e, g) and 1000\,G (panels b, d, f, h) vertical uniform magnetic fields. Panels a-d correspond to the electron beam models with the spectral power law index $\delta{}=$3, different colors and markers correspond to the different low-energy cutoffs, $E_{c}$. Panels e-h correspond to the electron beam models with $E_{c}=$25\,keV marked by different $\delta{}$. Black dashed horizontal lines mark uncertainties of the HMI observables at the disc center caused by the photon noise \citep{Couvidat16a}.}
	\label{figure4}
\end{figure}

	In this work we analyzed the available RADYN electron beam heating models in order to better understand the variations of the SDO/HMI observables (line-of-sight magnetic field, Doppler velocity, line depth, and continuum) during the impulsive phase of solar flares. We can draw the following conclusions from the study:
	
	\begin{enumerate}
		\item Because HMI observables are obtained from individual filtergrams distributed in time, the Fe\,I\,6173\,{\AA} line depth, Doppler velocity, and magnetic field strength measured by HMI during the flare impulsive phase can significantly deviate from the actual values. For beam heating events with average deposited energy fluxes of $F=5.0\times{}10^{10}$ erg\,cm$^{-2}$s$^{-1}$, the deviations can be as strong as 0.40\,km/s for the Doppler velocity and about 44\,G and 89\,G for the background vertical 100\,G and 1000\,G magnetic field.
		\item The strongest deviations of the velocity and the strongest decrease of the measured magnetic field depend on the background vertical magnetic field strength as demonstrated in Figure~\ref{figure3}.
		\item The deviations of the Doppler velocity and magnetic field observables most strongly correlate with the energy flux carried by non-thermal electrons at $\sim$50\,keV, while the continuum intensity enhancement is correlated most strongly with the $\sim$35\,keV electron flux. The correlation of the continuum intensity enhancement with the HXR photon flux also peak at $\sim$55\,keV.
		\item The current 1D radiative hydrodynamic flare models augmented with the uniform vertical magnetic field setup do not explain the strong magnetic field transients, sharp changes of the LOS velocities, and continuum enhancements observed during solar flares by the HMI instrument.
	\end{enumerate}
	
	There are several qualitative conclusions about possible misinterpretation of the HMI magnetic field measurements during solar flares that can be derived from this work. Although the artificial decrease of the HMI magnetic field observable can be strong (reaching almost 100\% for 50\,G and about 44\% for 100\,G as demonstrated in Figure~\ref{figure3}), the magnetic field observable still does not change its sign. Previously, magnetic field reversals were studied by \citet{Harker13a} who also did not reproduce this effect (Stokes V component reversal) by considering forward modeling of the Fe\,I\,6173\,{\AA} line and concluded that the sign reversal can be reached only if the Fe\,I\,6173\,{\AA} line profile goes to emission, which has not been observed. In our study, the modeled Fe\,I\,6173\,{\AA} line profiles are also always absorption profiles (i.e., the line depths derived from the exact line profile shapes are always positive). In addition, there is a clear tendency that it is harder to obtain significant relative magnetic field changes and sign reversal in stronger magnetic fields (see Figure~\ref{figure3}b) as reported by \citet{Kosovichev01a,Harker13a}. However, one has to take into account that the radiative hydrodynamic models utilized for this study did not have magnetic field evolution.
	
	The analysis of the continuum intensity enhancements also does not explain the results previously reported in the literature. For example, \citet{Prochazka18a} illustrated an average enhancement of SDO/HMI continuum intensity observable of more than 10\% in some sources during the 2014 June 11 X\,1.0 class flare. \citet{Song18c} performed a statistical analysis of the HMI continuum enhancement during the circular-ribbon flares of C, M, and X-class. The authors found that the magnitudes of continuum enhancement maxima, if observed, can be more than 10\% for C-class flares and can reach 100\% for the X-class flares. \citet{Macrae18a} reported the enhancement of the HMI continuum intensity integrated over the helioseismic source for about 7\% during the flare peak. In contrast, the continuum enhancements observed for the utilized SDO/HMI LOS observables for electron beam heating models (80 models utilized in our work) never exceed 3\%.
		
	Previously, \citet{Mravcova17a} concluded that the continuum enhancement observed by SDO/HMI may be an artifact due to simplified procedure of computing HMI observables. In the work by \citet{Svanda18a}, the authors considered joint observations of the X\,9.3 solar flare by SDO/HMI and Hinode Solar Optical Telescope Spectro-polarimeter \citep[Hinode SOT/SP,][]{Kosugi07a,Tsuneta08a}. Hinode/SOT spectropolarimetric data was inverted using the SIR code~\citep{RuizCobo92}, and the resulting atmospheric models were used to synthesize Fe\,6173\,{\AA} Stokes profiles, using the SIR code. Then the synthesized line continuum was compared with the corresponding HMI observable. The disagreement between the observed and modeled continua is found to be mostly within 10\% from modeled values \citep[see Figure 5 of][]{Svanda18a}, with some points deviating for 20\% or more. For the RHD flare models considered in this work, the disagreement between the measured and actual continuum intensity levels is of the same order as for the continuum enhancement, and never exceeds 3.2\%. This is, again, significantly lower than the values reposted by \citet{Svanda18a}. The authors also found the presence of synthesized Fe\,I\,6173\,{\AA} line profiles in emission that are not found in our simulations.
	
\begin{figure}[t!]
	\centering
	\includegraphics[width=1.0\linewidth]{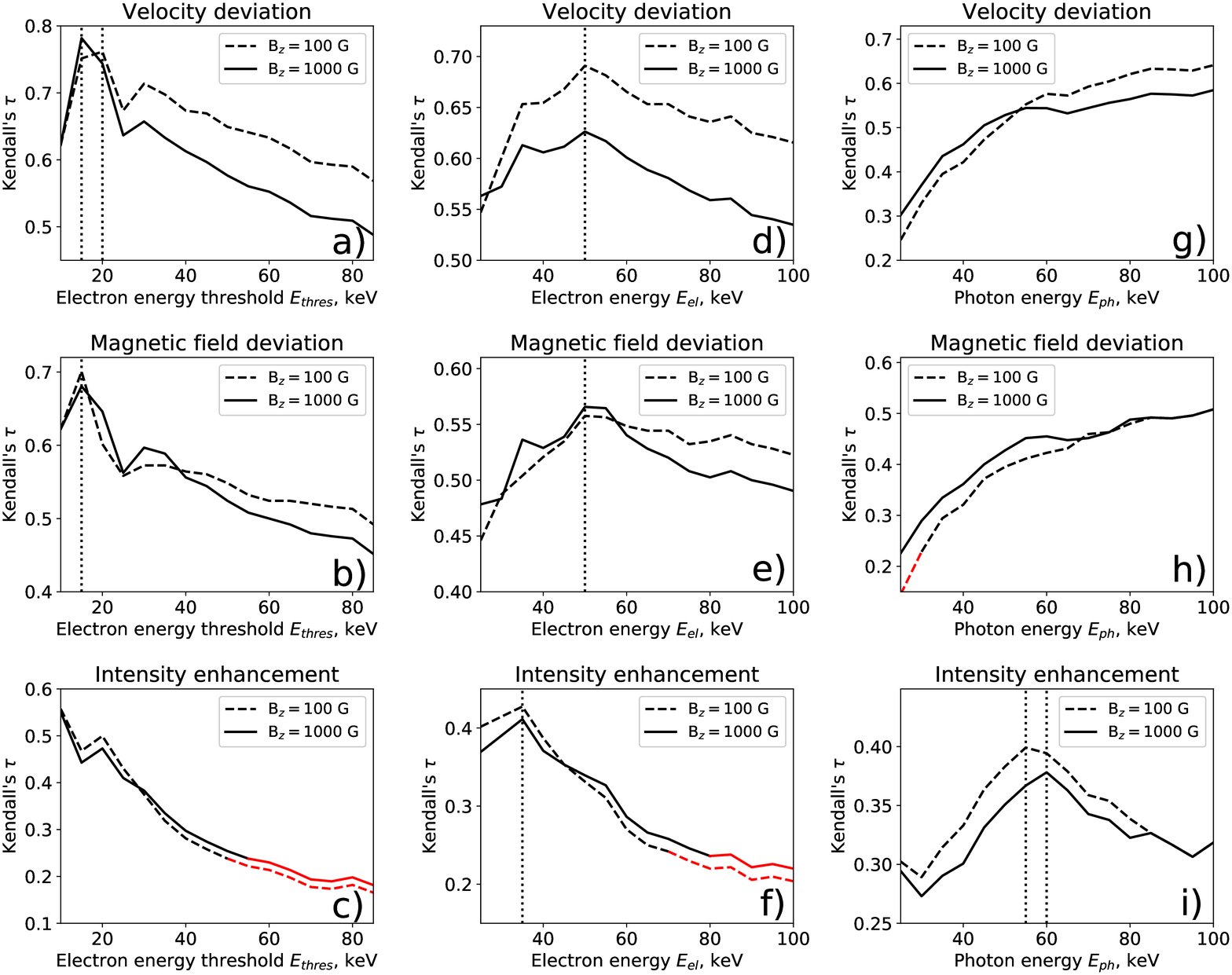}
	\caption{Kendall's $\tau$ correlation coefficients for the models with $F\ge{}1.5\times{}10^{10}$\,erg\,cm$^{-2}$s$^{-1}$ of the electron energy flux above the energy threshold, $F(E>E_{thres})$, and (a) the strongest deviations of LOS velocity, (b) the strongest deviations of magnetic field, or (c) the strongest enhancement of the continuum intensity. The corresponding correlation coefficients for the peak electron energy fluxes, $F(E=E_{el})$, and peak photon fluxes, $F_{ph}(E=E_{ph})$, are demonstrated in panels (d-f) and (g-i) correspondingly. Dashed lines correspond to the vertical uniform 100\,G field, solid lines~--- to 1000\,G field. The red part of the plots in panels (c), (f), and (h), indicates the energies where the p-value $>$ 0.05. The black dotted lines indicate the energies for which the strongest correlations are found.}
	\label{figure5}
\end{figure}

	Our study reveals that the strongest deviations of the derived observables are correlated not only with the deposited energy fluxes, but also with the energy fluxes carried by non-thermal electrons above a certain energy threshold, $F(E>E_{thres})$, and at a certain energy, $F(E=E_{el})$, as well as with the synthesized HXR photon flux of a certain energy, $F_{ph}(E=E_{ph})$. The correlations are statistically-significant (with respect to the assumption of absence of correlations) for almost any value of $E_{thres}$, $E_{el}$, and $E_{ph}$ (except high-energy $E_{thres}$ and $E_{el}$ for continuum intensity enhancements), and often peak at a certain energy value. In particular, Figure~\ref{figure5}i suggests that the continuum intensity enhancement most strongly correlates with the HXR photon flux at energies of $\sim$55-60\,keV. From observational perspectives, \citet{Huang16a} found that there was no obvious correlation between the 50\,keV photon flux and the equivalent area of a white-light flare. Contrary, based on the statistical analysis of 43 flares, \citet{Kuhar16a} found that the correlations between the HXR flux and the white light flux are the strongest for the HXR photons of $\sim$30-40\,keV energies. These energies are lower than those found for the models, however, both results are in agreement with the prediction of the HXR flux energy range which correlates most strongly with the white light enhancement. One also can notice in Figure~\ref{figure5} that the correlation coefficients for the continuum intensity enhancement (panels c, f, and i) are consistently lower than the correlation coefficients for the velocity and magnetic field deviations, especially for high energies. This gives an idea that, although high-energy electrons in the tail of the power law distribution can penetrate deep in the atmosphere and play an important role in the velocity and magnetic field perturbations, they are possibly less important for the observed continuum intensity enhancement.
	
	One of the restrictions of our study is that the initial atmospheric models were close to the quiet-Sun VAL3C atmospheres \citep{Vernazza81a}. However, \citet{Hong18a} demonstrated that, for the same electron beam heating setup, the Fe\,I\,6173\,{\AA} line profile experiences stronger perturbations if the initial atmosphere model represents the conditions of the sunspot penumbra rather than the quiet Sun. One more strong restriction is that the magnetic field variations are not taken into account consistently in RHD simulations. One has to consider flare models with evolving magnetic fields \citep{Cheung19a} to draw stronger conclusions. Nevertheless, the 3D radiative MHD simulations capable of modelling accurately the response of the photosphere to the beam heating events do not exist yet (i.e. the field-aligned RHD models still remain the state-of-the-art with that respect). We did not consider non-equilibrium effects on either the hydrogen populations (and therefore background opacity) nor on the Fe I populations. Recently \citet{Kerr19b,Kerr19c,Kerr19a} have investigated non-equilibrium radiation transfer during flares for Mg\,II and Si\,IV. They also studied the influence on the Mg\,II near-ultraviolet spectra of including non-equilibrium hydrogen populations (and opacity) when modeling these lines in flares. It was noted that, in the case of Mg\,II, up to 10\% intensity differences in the line wings can occur if non-equilibrium hydrogen populations are included, and that non-equilibrium effects on Mg II itself were non-negligible in the initial heating and decay phase of the flares. Since the hydrogen continuum can be important for the formation of Fe\,I, non-equilibrium effects of both hydrogen and Fe will be the focus of a further study. 
	
	The flare models have a certain imposed time dependence of the heating phase (triangular-shaped 20\,s heating) which is not necessarily the case for a particular solar flare. The actual transients may last significantly longer in time, up to several tens of minutes \citep{Sun17a,Castellanos18a}, and, obviously, require a separate explanation for their duration. Current loop models decay very quickly after the cessation of energy deposition, while observations show that flares cool over several tens of minutes or longer \citep[e.g.][]{Ryan13a}. This could be due to suppression of thermal conduction by non-local effects and turbulence \citep[e.g.][]{Emslie18a}, or due to post-impulsive phase energy deposition \citep[e.g.][]{Qiu16a}, both of which are actively being investigated using the RADYN code but beyond the scope of this present work. Further, we are modelling one flaring loop whereas an HMI pixel is an unresolved flare footpoint, and, perhaps, should be considered in the framework of multithread flare models \citep{RDC16a,Reep18a}. Despite the differences in timescale we can still make important contributions to understanding the physical mechanisms at play. We conclude that the current single-loop RHD flare models augmented with the uniform vertical magnetic field setups cannot explain the perturbations of the photospheric Fe\,I\,6173\,{\AA} line during solar flares. However, for correct interpretation of the SDO/HMI observables during solar flares, it is necessary to model line formation and variations of the line profile, taking into account the HMI observing sequence and data analysis procedure.

\acknowledgments

The research leading to these results has received funding from the European Community’s Seventh Framework Programme (FP7/2007-2013) under grant agreement no. 606862 (F-CHROMA). We also acknowledge the Stanford Solar Observatories Group and NASA Ames Research Center for the possibility to use the computational resources. GSK was funded by an appointment to the NASA Postdoctoral Program at Goddard Space Flight Center, administered by USRA through a contract with NASA. The research was partially supported by the NASA Grants: NNX12AD05A, NNX14AB68G, NNX16AP05H; and NSF grant 1639683.

\bibliographystyle{aasjournal}

\bibliography{HMIsim_RADYN}

\newpage

\newpage

\newpage

\newpage

\newpage

\end{document}